\documentclass[journal=jacsat,manuscript=paper,layout=twocolumn]{achemso}

\usepackage[version=3]{mhchem} 
\usepackage{bm}
\usepackage{physics}
\usepackage{color}
\usepackage{comment}
\usepackage[inkscapeformat=eps]{svg}
\usepackage{graphicx}

\newcommand{\rflct}{{\varrho}}

\usepackage{soul}
\usepackage{ifthen}
\newboolean{showannotations}   
\setboolean{showannotations}{true} 
\newcommand*{\added}[1]{%
  \ifthenelse{\boolean{showannotations}}{{\color{blue} #1}}{#1}%
}

\newcommand*{\removed}[1]{%
  \ifthenelse{\boolean{showannotations}}{{\color{red} \st{#1}}}{#1}%
}

\newcommand*{\change}[2]{%
  \ifthenelse{\boolean{showannotations}}{{\color{red}\st{#1}}{\color{blue}#2}}{#2}%
}

\author{\small Michael A.D. Taylor}
\affiliation[]
{Theoretical Division, Los Alamos National Laboratory, Los Alamos, NM, 87545.}
\email{mike_tay@lanl.gov}

\author{\small Jonathan Soderquist}
\affiliation[]
{The Institute of Optics, Hajim School of Engineering, University of Rochester, Rochester, New York 14627, USA}

\author{\small Pengfei Huo}
\email{pengfei.huo@rochester.edu}
\affiliation[]
{Department of Chemistry, University of Rochester, 120 Trustee Road, Rochester, New York 14627, USA}
\alsoaffiliation[]
{The Institute of Optics, Hajim School of Engineering, University of Rochester, Rochester, New York 14627, USA}
\alsoaffiliation[]
{Center for Coherence and Quantum Science, University of Rochester, Rochester, New York 14627, USA}

\title[]
  {\large Effective Mode Description for Macroscopic Fabry P\'{e}rot Cavities}

\abbreviations{IR,NMR,UV}
\keywords{American Chemical Society, \LaTeX}

\begin{document}

\begin{abstract}
{\small
We introduce an effective modes formalism to describe how the quasi-continuum of photonic modes in an optical cavity effectively behaves in the strong light-matter coupling regime of cavity quantum electrodynamics. By expressing these effective modes, we are able to reproduce the cavity dispersion relation while showing that the mode volumes of these effective modes are independent of the physical area of the Fabry-P\'erot cavity mirrors and are instead a measure of light-matter coupling strength, dependent on cavity design parameters such as mirror reflectivity.
\\
\\
\\
}

\begin{figure}
    \centering
    \includegraphics[width=1.0\linewidth]{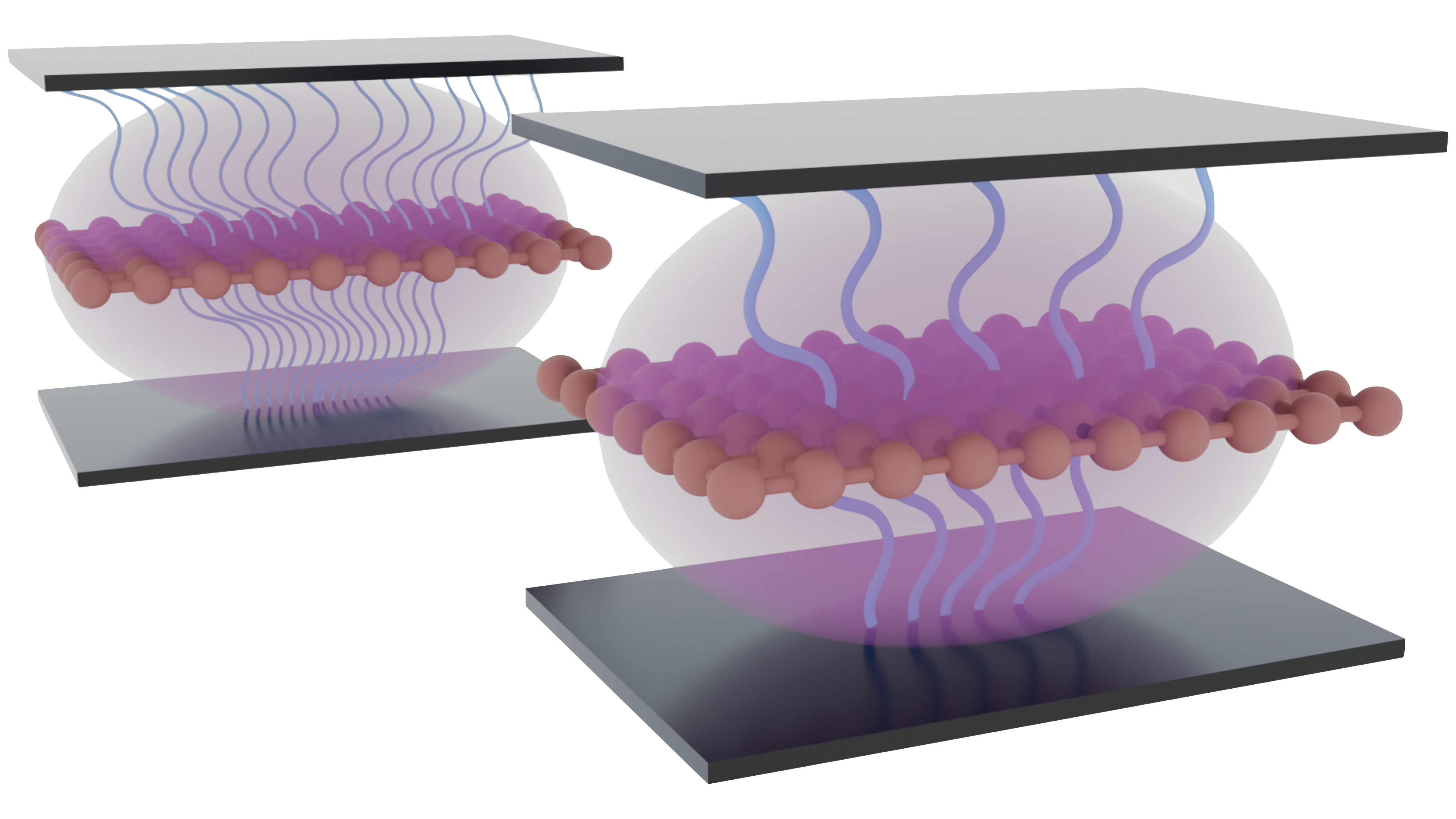}
\end{figure}

\end{abstract}
\maketitle

\section{\large Introduction}
{\small Quantum electrodynamics (QED) has been extremely successful in describing the fundamental quantum interaction between light and matter~\cite{CohenTannoudji1997}. Different applications of this theory, from quantum optics~\cite{Jaynes1963PotI,Mischuck,Hofheinz2008,Hofheinz2009} to polariton chemistry~\cite{Flick2017PNAS,Ebbesen2016ACR,Feist2018,Ribeiro2018,Mandal2023CR, Taylor2025CPR}, have been at the forefront of physics. Many approximations, including the two-level approximation, the rotating-wave approximation, and the neglect of second-order terms such as the dipole self-energy (in the multipolar gauge) or the diamagnetic term (in the Coulomb gauge) have historically been sufficient to replicate experimental results. However, in recent years, experimental advances in optical cavity design have produced light-matter coupling strengths for which these approximations are no longer valid~\cite{Bernardis2018PRA,Bernardis2018PRAa,Kockum2019NRP,Mandal2023CR, Taylor2025CPR}. This, in conjunction with the recent increase in computational power, has led to a revival of exact, fundamental forms of cavity QED~\cite{Stefano2019NP,Bernardis2018PRA, Taylor2020PRL,Stokes2022RMP}.

The most fundamental cavity quantum electrodynamics (QED) Hamiltonian~\cite{CohenTannoudji1997} is the minimal coupling Hamiltonian (also known as the ``p$\cdot$A'' Hamiltonian). However, for many experimentally realizable systems~\cite{Qiu2021JPCL,Li2021ARPC,Schwartz2013CPC,Thomas2016ACIE,Thomas2019S,Thomas2020N,Thomas2021,Vergauwe2019ACIE,Ebbesen2016ACR,GarciaVidal2021S,Hutchison2012ACIE,George2016PRL,Lather2019ACIE,Nagarajan2021JACS,Sau2021ACIE,Hirai2021CS,Hirai2021CL,Satapathy2021SA,Takele2021JPCC,Wiesehan2021JCP, Imperatore2021JCP,Vergauwe2019ACIE,Lather2022CS}, the immense number of photonic modes that are physically relevant makes even including all degrees of freedom (DOFs) as just two-level systems computationally intractable. Inevitably, the number of modes being considered must be numerically truncated~\cite{Taylor2022OL, Mandal2023CR, Taylor2025CPR} to perform meaningful calculations. 

In current literature, the few-mode approximation is commonly done by coarsely sampling the dispersion relation of the cavity~\cite{Mandal2023CR, Taylor2025CPR,Georgiou2021JCP, Balasubrahmaniyam2021PRB, Taylor2024PRB,Amin2024AN,Morshed2024JCP,Koessler2025,Mondal2023JCP,Weight2023PCCP,Weight2024c,Chng2025NL}. By doing this, we obfuscate the source of the light-matter coupling strength ($\propto 1/\sqrt{V}$) as the quantization volume, $V$, becomes difficult to identify~\cite{Svendsen2025CP}. As such, we tend to treat the coupling strength as an experimentally measured parameter instead of a physically defined property of the polaritonic system. For strong coupling strengths, the corresponding quantization volume required is microscopic in scale. However, we know that for polariton experiments using Fabry-P\'erot cavities, the geometric volume of the cavity is at least mesoscopic in scale~\cite{Qiu2021JPCL,Li2021ARPC,Schwartz2013CPC,Thomas2016ACIE,Thomas2019S,Thomas2020N,Thomas2021,Vergauwe2019ACIE,Ebbesen2016ACR,GarciaVidal2021S,Hutchison2012ACIE,George2016PRL,Lather2019ACIE,Nagarajan2021JACS,Sau2021ACIE,Hirai2021CS,Hirai2021CL,Satapathy2021SA,Takele2021JPCC,Wiesehan2021JCP, Imperatore2021JCP,Vergauwe2019ACIE,Lather2022CS,Amin2024AN,Morshed2024JCP}. This creates a mystery as to how cavities with mesoscopic and even macroscopic mirrors can generate strong light-matter coupling. 

Similarly, from basic intuition, one can deduce that there should be some relation between the quality factor of the cavity and the coupling strength of the confined field, since in the limit of the quality factor going to zero, the coupling strength should approach the vacuum coupling strength. Experimentally, it has also been verified that coupling strength is directly related to the field intensity of a cavity mode\cite{Ahn2018ACSPh}. However, in the majority of works that even consider cavity loss, the coupling strength and the cavity quality factor are treated as orthogonal parameters~\cite{Amin2024AN,Morshed2024JCP,Koessler2025,Mondal2023JCP,Weight2024c,Chng2025NL,Ying2024JCP,Mondal2025JCP,Mondal2025JCPa}. In these cases, the ideal cavity dispersion is sampled with the cavity loss introduced phenomenologically through Lindbladian terms and broadening factors in the absorption spectra.

In their recent work,~\cite{Svendsen2025CP} Svendsen et.al. took a significant step towards developing a simple, effective model of the cavity to better understand quantities like the effective mode volume by integrating over the entire relevant energetic range of the cavity dispersion to form a single strongly coupled mode with the average energy of the integrated range. However, due to the nature of this model, it no longer accounts for the cavity dispersion relation's frequency dependence, the conservation of momentum between the light and matter degrees of freedom, and the cavity quality factor. 

In this paper, we build off of Ref.~\citenum{Svendsen2025CP} to better understand how the quasi-continuum of photonic modes inside an optical cavity fundamentally behaves in the strong light-matter coupling regime. To do this, we formulate a new effective modes framework to encapsulate the effects from all the photonic modes in a small finite number of effective modes, enabling us to address the questions of cavity quantization volume and the relation between cavity coupling and loss. We start our analysis with the most fundamental QED Hamiltonian, not subject to any approximations, and deliberately layer on approximations to create a framework that is computable. However, this work's impact is beyond reformulating the Hamiltonian for computational advantage. By expressing these effective modes, we can solve current mysteries in the field, leveraging the formalism and intuition gained in this framework. Namely, we can derive the effective mode volumes for modes inside realistic, lossy cavities. Further, we show that this mode volume is independent of the area of the cavity mirrors.

The remainder of this paper is structured in four more sections. We begin by reviewing the second quantization on the exact Coulomb gauge Hamiltonian, demonstrating explicitly the conservation of momentum between the matter and photonic DOFs. Next, we then use an ideal Fabry-P\'erot cavity model to introduce the effective modes framework, demonstrating the basis reduction while still considering the physics from all modes. Then, we expand the formalism to fully describe a realistic, lossy Fabry-P\'erot cavity, allowing us to derive the mode quantization volume for each effective mode. Finally, we summarize and conclude our analysis.

\section{\normalsize Conservation of Momentum Between Light and Matter} \label{sec:ConMom}

We begin our analysis with the most fundamental QED Hamiltonian~\cite{CohenTannoudji1997}, the minimal coupling Hamiltonian (also known as the ``p$\cdot$A'' Hamiltonian) in the Coulomb gauge ($\nabla\cdot {\bf A}=0$).
\begin{align}\label{eq:p.A}
    \hat{H}_\mathrm{p\cdot A} =&~ \sum_i^{N_e} \frac{1}{2m} \Big( \hat{\bf p}_i - e\hat{\bf A}(\hat{\bf r}_i) \Big)^2 + \frac{1}{2} \sum_{i \neq j}^{N_e} v_\mathrm{ee}(\hat{r}_i,\hat{r}_j)  \nonumber \\
    &+ \sum_i^{N_e} v_\mathrm{eN}(\hat{r}_i) + \sum_{\bf{q},\lambda} \hbar \omega_{\bf{q}} \Big(\hat{a}_{\bf{q},\lambda}^\dagger \hat{a}_{\bf{q},\lambda} + \frac{1}{2} \Big)
\end{align}
where $\{i,j\}$ index over the $N_e$ electrons (of charge $e$) in the system with the position of $\hat{\bf r}_i$ and canonical momentum of $\hat{\bf p}_i$, which experience a two-body Coulomb potential, $v_\mathrm{ee}(\hat{r}_i,\hat{r}_j)$, and a single-body Coulomb potential with nuclei, $v_\mathrm{eN}(\hat{r}_i)$. $\{ \bf q, \lambda\}$ index over the photonic modes of wavevector, $\bf q$, and polarization $\lambda$, whose creation (annihilation) operators are $\hat{a}_{\bf{q},\lambda}$ ($\hat{a}_{\bf{q},\lambda}^\dagger$). Note that we made no long wavelength approximation, so the vector potential of the EM field, $\hat{\bf A} ( {\bf r} )$, can be written as
\begin{equation}\label{eq:A}
    \hat{\bf A}({\bf r}) = \sum_{{\bf q},\lambda}  e^{i {\bf q}\cdot {\bf r}} \hat{\boldsymbol{\epsilon}}_{{\bf q},\lambda} e\sqrt{\frac{\hbar}{ 2\varepsilon_{0} \omega_{\bf q} {V}}}   \big( \hat{a}_{{\bf q},\lambda} + \hat{a}^{\dagger}_{-{\bf q},\lambda}\big),
\end{equation}
with explicit spatial dependence. Going forward we set the shorthand ${A}_{\bf q} = e\sqrt{\frac{\hbar}{ 2\varepsilon_{0} \omega_{\bf q}}}$, the quantization volume-independent component of the vector potential magnitude.

To better understand the effect of the spatial dependence of $\hat{\bf A}(\hat{\bf r}_i)$, we will also perform the second quantization on the matter degrees of freedom. For the purposes of this paper, we assume that the matter coupled to the cavity is a 2D material such that it is periodic in the x-y plane (parallel to the mirrors) but not in the z-direction (perpendicular to the mirrors) that is located equidistant from the mirrors ($z=0$). We define the annihilation operator, $\hat{\psi}({\bf{r}_\|}, \sigma)$, using Bloch's Theorem as
\begin{align}\label{eq:psi}
    \hat{\psi}({\bf{r}_\|}, \sigma) &= \frac{1}{\sqrt{S_\mathrm{M}}} \sum_{j, {\bf k}_\|, \sigma}^\mathrm{BZ} e^{i {\bf k}_\| \cdot {\bf{r}_\|}} \, u_{j, {\bf k}_\|} ({\bf r}_\|, \sigma) \, \hat{c}_{j, {\bf k}_\|, \sigma} 
\end{align}
where $\{ {\bf{r}_\|}, \sigma\}$ are the in-plane spatial coordinate and the spin coordinate for $\hat{\psi}$, respectively, and $\{{\bf k}_\|, j\}$ index over the in-plane electronic wavevector and band within the first Brillouin zone, respectively. $u({\bf r}_\|, \sigma)=u({{\bf r}_\|+{\bf R}_n}, \sigma)$ is the Bloch function for the $j_\mathrm{th}$ band, where ${\bf R}_n$ is the coordinate of the $n_\mathrm{th}$ site. $S_M$ is the area of the crystal unit cell in the x-y plane, and $\hat{c}_{j, {\bf k}_\|, \sigma}$ is the electronic annihilation operator. By casting Eq.~\ref{eq:p.A} into the basis of {$\hat{\psi}({\bf{r}_\|}, \sigma)$}, we can now write the $\mathrm{p\cdot A}$ Hamiltonian under second quantization for both light and matter as
\begin{align} \label{eq:p.A-SQ}
\hat{H}_\mathrm{p\cdot A}  = &\hat{H}_\mathrm{el} + \hat{H}_\mathrm{ph} \\
&- \sqrt{\frac{1}{V}} \sum_{i,j, {\bf k}_\|, \sigma} \, \sum_{{\bf q},\lambda} ~~ \hat{c}_{i, {\bf k}_\| + {\bf q}_\|, \sigma}^\dagger \hat{c}_{j, {\bf k}_\|, \sigma} \, \nonumber\\
&~~~~\times A_{\bf q}  \, {\bf p}_{i,j,{\bf k}_\|,{\bf q}}(z) \cdot \hat{\boldsymbol{\epsilon}}_{{\bf q},\lambda} \big( \hat{a}_{{\bf q},\lambda} + \hat{a}^{\dagger}_{-{\bf q},\lambda} \big) \nonumber \\
&+ \frac{1}{2V} \sum_{i,j, {\bf k}_\|, \sigma} \, \sum_{{\bf q},\lambda,{\bf q}',\lambda'}  \,  \hat{c}_{i, {\bf k}_\| + {\bf q}_\| + {\bf q}_\|', \sigma}^\dagger \hat{c}_{j, {\bf k}_\|, \sigma}    \nonumber \\
    &~~~~\times s_{i,j,\sigma, {\bf k}_\|,{\bf q},{\bf q}'}(z) \, A_{\bf q} \, A_{\bf q'} \hat{\boldsymbol{\epsilon}}_{{\bf q},\lambda} \cdot \hat{\boldsymbol{\epsilon}}_{{\bf q}',\lambda'} \nonumber \\
    &~~~~\times\big( \hat{a}_{{\bf q},\lambda} + \hat{a}^{\dagger}_{-{\bf q},\lambda} \big) \big( \hat{a}_{{\bf q}',\lambda'} + \hat{a}^{\dagger}_{-{\bf q}',\lambda'} \big)\nonumber,
\end{align}
where $\hat{H}_\mathrm{ph} = \sum_{\bf{q},\lambda} \hbar \omega_{\bf{q}} (\hat{a}_{\bf{q},\lambda}^\dagger \hat{a}_{\bf{q},\lambda} + \frac{1}{2} )$ and
\begin{align}
    &{\bf p}_{i,j,{\bf k}_\|,{\bf q}}(z) = \int_{\Omega_\|} d{r} \,\, u_{i, {\bf k}_\| + {\bf q}_\|}^* ({\bf r}, \sigma) (-i \hbar {\nabla} + \hbar {\bf k}_\|)\,  \nonumber \\
    &~~~~~~~~~~~~~~~~~~~~~~~~~~\times u_{j, {\bf k}_\|}^* ({\bf r}, \sigma) \\
    &s_{i,j,\sigma, {\bf k}_\|,{\bf q},{\bf q}'} (z) = \int_{\Omega_\|} d{r} \,\, u_{i, {\bf k}_\| + {\bf q}_\| + {\bf q}'_\|}^* ({\bf r}, \sigma) \, u_{j, {\bf k}_\|}^* ({\bf r}, \sigma)
\end{align}
are the matrix elements for the momentum operator and wavefunction overlaps, respectively, for $\Omega_\|$ being a unit cell. By expressing the p$\cdot$A Hamiltonian under second quantization, the conservation of momentum between light and matter becomes immediately apparent.~\cite{Li2024JCP,Li2024JCTC,Svendsen2025CP} If a photon of in-plane momentum, $\bf q_\|$, is created through $\hat{a}^{\dagger}_{{\bf q},\lambda}$, then that momentum must be removed from the matter through the term, $\hat{c}_{i, {\bf k}_\| - {\bf q}_\|, \sigma}^\dagger \hat{c}_{j, {\bf k}_\|, \sigma}$. Similarly, for the two-photon diamagnetic term (second line of Eq.~\ref{eq:p.A-SQ}), the matter DOFs must accommodate the net momentum change from the two photons.

With this rigorous, second-quantized Hamiltonian, we begin to see the inherent basis scaling difficulty with even just the photonic DOFs. As shown explicitly with the conservation of momentum, there are all-to-all second-order interactions between the quasi-continuous spectrum of photonic modes, mediated through the momentum exchange with matter. For even the simplest eigen-energy calculation with the simplest model matter Hamiltonian, this problem becomes intractably expensive. As such, any calculation must reduce the number of bosonic modes considered.

\section{\normalsize Effective Mode Description for Ideal Cavity} \label{sec:Ideal}

We begin our effective modes analysis with a simple model cavity. We assume a Fabry-P\'{e}rot cavity with perfect mirrors, separated a distance of $L_\mathrm{c}$, and truncate the modes in the $z$-direction (perpendicular to the mirrors) to only consider the $m_\mathrm{th}$ order cavity modes, where $|q_z| = \frac{m\pi}{ L_\mathrm{c}}$. The wavevector components parallel to the mirrors are quasi-continuous, with the difference in $\bf q_\|$ along either axis being $d{q}_\| = d{q}_x = d{q}_y = 2 \pi / L$, with $L$ being the length of the mirrors (assumed to be equal in $x$- and $y$- directions for simplicity). As $L$ is typically on the millimeter scale, $d{q}_\|$ is extremely small. Due to the exact boundary conditions in the $z$-direction, the cavity has a 2D dispersion relation of
\begin{equation} \label{eq:ideal_disp}
    \omega = \frac{c}{n_\mathrm{r}} \sqrt{q_x^2 + q_y^2 + \bigg( \frac{m\pi}{ L_\mathrm{c}}\bigg)^2},
\end{equation}
where $n_\mathrm{r}$ is the refractive index inside the cavity, such that any mode on the light line that does not satisfy the above equality is forbidden inside the cavity. Going forward, we define the component of the wavevector in the $xy$-plane as $\bf q_\|$.

In practice, modeling the quasi-continuous number of modes to perform even a simple eigenenergy calculation is computationally intractable due to the exponential scaling of the basis size with the number of modes. Typically, the photonic dispersion is coarsely sampled for a small number of modes ($\sim$15-20), ignoring the other modes that contribute to the light-matter coupling.

Instead of sampling, we will divide the dispersion into a Cartesian grid of $N$ bins with the width of each bin, $\Delta {q}_\|= \Delta {q}_x = \Delta {q}_y$, being a convergence parameter. The wavevectors and frequencies for all modes within each bin are approximated as equal. In contrast to the coarse sampling method, this allows every mode to still contribute to the light-matter coupling. We can then write the photonic Hamiltonian in this manner as
\begin{align}
    \hat{H}_\mathrm{ph} &= \sum_{n,\lambda} \sum_{{q}_x = {q}_{n,x} - \Delta {q}_{\|}}^{{q}_{n,x} + \Delta {q}_{\|}} \sum_{{q}_y = {q}_{n,y} - \Delta {q}_{\|}}^{{q}_{n,y} + \Delta {q}_{\|}} \hbar \omega_{\bf q} \bigg( \hat{a}_{\bf q}^\dagger \hat{a}_{\bf q} + \frac{1}{2} \bigg) \nonumber \\
    &\approx \sum_n \hbar \omega_{{\bf q}_n} \sum_{{q}_x = {q}_{n,x} - \Delta {q}_{\|}}^{{q}_{n,x} + \Delta {q}_{\|}} \sum_{{q}_y = {q}_{n,y} - \Delta {q}_{\|}}^{{q}_{n,y} + \Delta {q}_{\|}} \bigg( \hat{a}_{\bf q}^\dagger \hat{a}_{\bf q} + \frac{1}{2} \bigg),
\end{align}
where ${\bf q}_{n}$ is the average wavevector for the $n_\mathrm{th}$ bin and the ${\bf q}$ inside the sums is the vector $[{q}_{x},{q}_{y},{q}_{z}]$. In the second line, we approximate the frequencies for all the modes within the bin to be $\omega_{{\bf q}_n} \equiv c |{\bf q}_{n}|/n_\mathrm{r}$. For simplicity of notation, from here on we will express $\sum_n \sum_{{q}_x = {q}_{n,x} - \Delta {q}_{\|}}^{{q}_{n,x} + \Delta {q}_{\|}} \sum_{{q}_y = {q}_{n,y} - \Delta {q}_{\|}}^{{q}_{n,y} + \Delta {q}_{\|}}$ as $\sum_n \sum_{{\bf q} \in n}$ with the understanding that the second sum is over all modes within the $n_\mathrm{th}$ bin. Now we have approximated the photonic Hamiltonian as $N$ bins of photonic modes, each comprised of 
\begin{equation}
\ell = (\Delta q_\| / dq_\|)^2
\end{equation}
nearly identical modes. Under the continuous mode limit (without coarse graining), $\Delta q_\| \sim dq_\|\to 0$ while $\Delta q_\| / dq_\|\to 1$, such that $\ell\to 1$. 

We then make one further approximation that the light-matter momentum exchange is identical for all modes within a given bin. As such, we can approximate Eq.~\ref{eq:p.A-SQ} as
\begin{align} \label{eq:p.A-ManyIdenModes}
    \hat{H}_\mathrm{p\cdot A} \approx&~ \hat{H}_\mathrm{el} + \sum_n \hbar \omega_{{\bf q}_n} \sum_{{\bf q} \in n} \bigg( \hat{a}_{\bf q}^\dagger \hat{a}_{\bf q} + \frac{1}{2} \bigg) \\
    &- \sqrt{\frac{1}{V}} \sum_{i,j, {\bf k}_\|, \sigma} \, \sum_{n,\lambda} \hat{c}_{i, {\bf k}_\| + {\bf q}_n, \sigma}^\dagger \hat{c}_{j, {\bf k}_\|, \sigma}  \nonumber \\
    &~~~~\times A_{{\bf q}_n}  \, \bar{p}_{i,j,{\bf k}_\|,{\bf q}_n,\lambda} \sum_{{\bf q} \in n}  \big( \hat{a}_{{\bf q},\lambda} + \hat{a}^{\dagger}_{-{\bf q},\lambda} \big) \nonumber \\
    &+ \frac{1}{2V} \sum_{i,j, {\bf k}_\|, \sigma} \, \sum_{n,\lambda,n',\lambda'} \hat{c}_{i, {\bf k}_\| + {\bf q}_n + {\bf q}_n', \sigma}^\dagger \hat{c}_{j, {\bf k}_\|, \sigma}   \nonumber \\
    &~~~~\times\bar{s}_{i,j,\sigma, {\bf k}_\|,{\bf q}_n,{\bf q}_n'} \, A_{{\bf q}_n} \, A_{{\bf q}_n'} \hat{\boldsymbol{\epsilon}}_{{\bf q}_n,\lambda} \cdot \hat{\boldsymbol{\epsilon}}_{{\bf q}_n',\lambda'} \nonumber \\ 
    &~~~~\times \sum_{{\bf q} \in n} \sum_{{\bf q}' \in n'} \big( \hat{a}_{{\bf q},\lambda} + \hat{a}^{\dagger}_{-{\bf q},\lambda} \big) \big( \hat{a}_{{\bf q}',\lambda'} + \hat{a}^{\dagger}_{-{\bf q}',\lambda'} \big), \nonumber
\end{align}
where we have now applied all the approximations previously mentioned to the full Hamiltonian. However, we have not gained any computational advantage yet with this, since there is no reduction in the number of photonic modes. 

Since each bin contains many identical modes that all couple to matter in an identical fashion, we can perform a normal mode transformation with a single effective mode carrying all of the interactions for each bin. To do so, we first express all of the modes in the $n_\mathrm{th}$ bin in their harmonic oscillator analogous coordinate ($\hat{x}_{{\bf q},\lambda}$) and momentum ($\hat{p}_{{\bf q},\lambda}$) operators. Then we can define the ``center-of-mass" collective coordinate with its conjugate momentum as well as the relative coordinates and momenta as
\begin{subequations}
\begin{gather}
    \hat{X}_{n, \lambda} \equiv \sqrt{\frac{\hbar}{2 \omega_n}} \sum_{{\bf q} \in n} \big( \hat{a}_{{\bf q},\lambda} + \hat{a}^{\dagger}_{-{\bf q},\lambda} \big) = \sum_{{\bf q} \in n} \hat{x}_{{\bf q},\lambda} \\
    \hat{\chi}_{n,{\bf q},\lambda} \equiv \hat{x}_{{\bf q},\lambda} - \frac{1}{\ell} \hat{X}_{n, \lambda} \\
    \hat{P}_{n, \lambda} \equiv - i \frac{\partial}{\partial \hat{X}_{n, \lambda}}\\
    \hat{\varphi}_{n,{\bf q},\lambda} \equiv \hat{p}_{{\bf q},\lambda} - \hat{P}_{n, \lambda}
\end{gather}
\end{subequations}
where $\hat{X}_{n, \lambda}$ and $\hat{P}_{n, \lambda}$ are the collective coordinate and momenta operators for the $n_\mathrm{th}$ bin with polarization $\lambda$, and $\hat{\chi}_{n,{\bf q},\lambda}$ and $\hat{\varphi}_{n,{\bf q},\lambda}$ are the relative coordinate and momenta operators for the same bin but indexed by the ${\bf q}_\mathrm{th}$ mode in the bin. Note that $\sum_{{\bf q} \in n} \hat{\chi}_{n,{\bf q},\lambda} = \sum_{{\bf q} \in n} \hat{\varphi}_{n,{\bf q},\lambda} = 0$.

We can then write the effective photonic Hamiltonian as
\begin{align} \label{eq:h_eff_ph_ideal}
    &\hat{H}_\mathrm{ph}^\mathrm{eff} = \sum_{n,\lambda} \sum_{{\bf q} \in n} \frac{1}{2} \big(\hat{p}_{{\bf q},\lambda}\hat{p}_{-{\bf q},\lambda} + \omega_n^2 \hat{x}_{{\bf q},\lambda}\hat{x}_{-{\bf q},\lambda} \big) \nonumber\\
    &= \sum_{n,\lambda} \frac{1}{2} \big( \ell \hat{P}_{n, \lambda}^2 + \frac{1}{\ell} \omega_n^2 \hat{X}_{n, \lambda}^2 \big) \\
    &~~+ \sum_{n,\lambda,{\bf q} \in n} \frac{1}{2} \big( \hat{\varphi}_{n,{\bf q},\lambda}\hat{\varphi}_{n,-{\bf q},\lambda}+ \omega_n^2 \hat{\chi}_{n,{\bf q},\lambda}\hat{\chi}_{n,-{\bf q},\lambda} \big), \nonumber
\end{align}
where the last term is the Hamiltonian for all the relative modes, which is completely decoupled from the matter system and the other modes. This means that the collective mode for each bin contains all of the light-matter coupling for all modes in the bin and can be treated as the \textit{effective} mode of the bin. Additionally, we can define the creation and annihilation operators for this collective mode using
\begin{subequations}
\begin{align}
    \hat{X}_{n, \lambda} &= \sqrt{\frac{\hbar \ell}{2 \omega_n}} \big(\hat{b}_n^\dagger +\hat{b}_n \big) \\
    \hat{P}_{n, \lambda} &= i \sqrt{\frac{\hbar \omega_n}{2 \ell}} \big(\hat{b}_n^\dagger - \hat{b}_n \big).
\end{align}
\end{subequations}
We can further compress our notation by defining the matter operators
\begin{subequations}
\begin{align}
    \hat{\bar{p}}_{n, \lambda} &= \sum_{i,j, {\bf k}_\|, \sigma} \hat{c}_{i, {\bf k}_\| + {\bf q}_n, \sigma}^\dagger \hat{c}_{j, {\bf k}_\|, \sigma} \, \bar{p}_{i,j,{\bf k}_\|,{\bf q}_n,\lambda} \, \\
    \hat{\bar{s}}_{n, n'} &= \sum_{i,j, {\bf k}_\|, \sigma} \hat{c}_{i, {\bf k}_\| + {\bf q}_n + {\bf q}_n', \sigma}^\dagger \hat{c}_{j, {\bf k}_\|, \sigma}   \, \bar{s}_{i,j,\sigma, {\bf k}_\|,{\bf q}_n,{\bf q}_n'},
\end{align}
\end{subequations}
allowing us to then more succinctly write the effective mode p$\cdot$A Hamiltonian as
\begin{align} \label{eq:p.A-eff}
    \hat{H}_\mathrm{p\cdot A} =&~ \hat{H}_\mathrm{el} + \sum_{n,\lambda} \hbar \omega_{n} \bigg( \hat{b}_{n,\lambda}^\dagger \hat{b}_{n, \lambda} + \frac{1}{2} \bigg) + \hat{H}_\mathrm{ph}^\mathrm{rel} \nonumber\\
    &+ \sqrt{\frac{\ell}{{V}}} \sum_{n,\lambda} \, A_{{\bf q}_n}  \hat{\bar{p}}_{n, \lambda} \big( \hat{b}_{n,\lambda} + \hat{b}^{\dagger}_{n,\lambda} \big) \nonumber \\
    &+ \frac{\ell}{2V} \sum_{n,\lambda,n',\lambda'} \hat{\bar{s}}_{n, n'} A_{n} A_{n'} \hat{\boldsymbol{\epsilon}}_{n,\lambda} \cdot \hat{\boldsymbol{\epsilon}}_{n',\lambda'}  \nonumber \\
    &~~~~\times\big( \hat{b}_{n,\lambda} + \hat{b}^{\dagger}_{n,\lambda} \big) \big( \hat{b}_{n',\lambda'} + \hat{b}^{\dagger}_{n',\lambda'} \big), 
\end{align}
where $\hat{H}_\mathrm{ph}^\mathrm{rel} =\sum_{n,\lambda,{\bf q} \in n} \frac{1}{2}  (\hat{\varphi}_{n,{\bf q},\lambda}^2 + \omega_n^2 \hat{\chi}_{n,{\bf q},\lambda}^2)$ is now completely decoupled from the rest of the Hamiltonian. As such, this portion can be ignored from calculations, allowing one to only need $N$ effective modes to calculate the energy eigenspectrum.

Additionally, one should consider the factor of $\ell / V$ present in both the p$\cdot$A and diamagnetic terms in Eq.~\ref{eq:p.A-eff}. Since $V = L^2 L_c$ for $L$ being the mirror length and $L_c$ being the distance between mirrors and $\ell = \Delta q_\|^2 / dq_\|^2 = \frac{\Delta q_\|^2 L^2}{4 \pi^2}$, $\ell / V$ is completely \textit{independent} of the mirror size and can be thought of as an effective cavity quantization volume for the effective mode, allowing us to define the effective mode volume, $V_\mathrm{eff} \equiv V / \ell$, as
\begin{equation}
    V_\mathrm{eff} = \frac{4 \pi^2 L_c}{  \Delta q_\|^2}.
\end{equation}
This result has a number of salient consequences. Computationally, it provides an accurate way to model only $N$ modes, yet still contains the full contributions of many modes. Theoretically, it begins to demonstrate how to understand the effective mode volume inside a cavity. However, this ideal cavity analysis relies on an ad-hoc choice of $\Delta q_\|$. To provide a more physical insight into how to form the bins, we must extend our formalism to consider a realistic, lossy cavity.


\section{\normalsize Effective Mode Description for a Lossy Cavity} \label{sec:Lossy}
While ideal cavities have a simple 2D dispersion relation as described in Eq.~\ref{eq:ideal_disp}, lossy cavities no longer follow this relation as $q_z$ is no longer exactly fixed. Instead, the cavity dispersion is the 3D dispersion of real space, $\omega_{\bf q} = {|\bf q|}{c} / n_\mathrm{r}$, and the effect of the cavity comes from an enhancement, $R_{\bf q},$ of the amplitude of the vector potential of modes inside the cavity. The vector potential in the cavity is given by
\begin{equation}
    \hat{\bf A}({\bf r}) = \sum_{{\bf q},\lambda}   \hat{\boldsymbol{\epsilon}}_{{\bf q},\lambda} \sqrt{\frac{\hbar}{ 2\varepsilon_{0} \omega_{\bf q} {V}}}   \big( R_{\bf q}\hat{a}_{{\bf q},\lambda}e^{i {\bf q}\cdot {\bf r}} + \mathrm{h.c.} \big)
\end{equation}
}

\begin{figure}[t]
    \centering
    \includegraphics[width=1.0\linewidth]{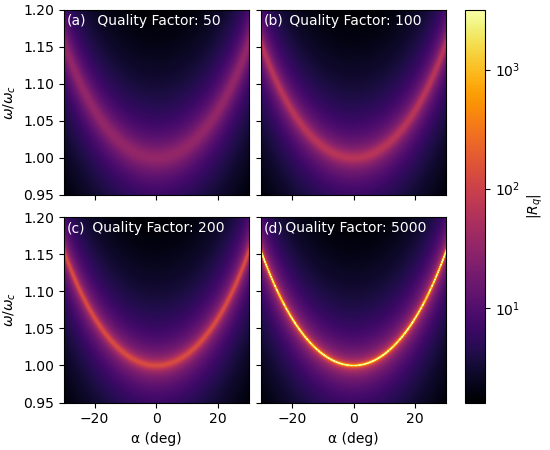}
    \caption{ \small Enhancement factor, $R_{\bf q}$, as a function of $\omega$ and $\alpha = \tan^{-1}(c q_\| / \omega_\mathrm{c})$ for {\bf (a)} ${\mathcal Q} = 50$, {\bf (b)} 100, {\bf (c)} 200, and {\bf (d)} 5000.}
    \label{fig:dispersion}
\end{figure}

For a Fabry-Perot cavity, with two identical mirrors, the enhancement function is
\begin{equation} \label{eq:R_q_z}
    R_{\bf q}(z) = \frac{1 + |\rflct| e^{i (|q_z| L_\mathrm{c} +\phi- 2{\bf q}_z \cdot {\bf z})}}{1 - |\rflct|^2 e^{2i (|q_z| L_\mathrm{c}+\phi)} },
\end{equation}
where $\rflct=|\rflct|e^{i\phi}$ is the complex-valued reflection coefficient of the cavity mirrors, and $q_z$ is approximated to be fully real inside the cavity. (See Supporting Information for a detailed derivation of the enhancement in a Fabry-Perot cavity along with useful approximations.) Vacuum fluctuations allow the $\bf q_z$ to be in the positive or negative $z$-direction, so both cases need to be taken into account as separate modes. Equation \ref{eq:R_q_z} also lends itself nicely to a coordinate system that contains $q_z$, so we choose to use the coordinate system $\{q_z, \omega, \theta\}$ throughout the rest of the paper, where $\theta \equiv \arctan(q_y / q_x)$ is the angle of propagation in the x-y plane. In Fig.~\ref{fig:dispersion}a-d, we plot $|R_{\bf q}(0)|$ for different values of $\rflct$ (a given quality factor, ${\mathcal Q}$, tuned to be resonant at normal incidence, corresponds to $\rflct$ via 
\begin{equation}
|\rflct| \approx e^{- \frac{m\pi}{2{\mathcal Q}}}\approx 1-\frac{m\pi}{2{\mathcal Q}},
\end{equation}
for ${\mathcal Q} \gg 1$ where $m$ is the mode order that the cavity is tuned to). Note that as ${\mathcal Q}$ increases, $|R_{\bf q}(0)|$ peaks more sharply near the ideal cavity dispersion. (See Supporting Information for a detailed discussion of the lossless cavity limit.)

{Using this mode enhancement factor, $R_{\bf q}(z)$ for a 2D material at $z=0$, we can write the vector potential for a lossy cavity as}
\begin{align}
    \hat{\bf A}({\bf r}) =  \frac{1}{\sqrt{V}}\sum_{{\bf q},\lambda}  e^{i {\bf q}_\|\cdot {\bf r}_\|} \hat{\boldsymbol{\epsilon}}_{{\bf q},\lambda} A_{\bf q}   \big( R_{\bf{q}}(0) \,\, \hat{a}_{q_z,{\bf q}_\|,\lambda} \nonumber\\
    +  R_{\bf{q}}^*(0) \,\, \hat{a}_{q_z,-{\bf q}_\|,\lambda}^\dagger\big).
\end{align}

{By performing a phase rotation for the photonic modes and defining {$R_{\bf q_z, \omega = c|\bf q|} \equiv |R_{\bf{q}}(0)|$}, we can rewrite $\hat{\bf A}({\bf r})$ in a more convenient form,}
\begin{equation}
    \hat{\bf A}({\bf r}) =  \frac{1}{\sqrt{V}}\sum_{{\bf q},\lambda}  e^{i {\bf q}_\|\cdot {\bf r}_\|} \hat{\boldsymbol{\epsilon}}_{{\bf q},\lambda} A_{\bf q} R_{q_z, \omega}  \big( \hat{b}_{q_z,-{\bf q}_\|,\lambda}^\dagger +   \,\, \hat{b}_{q_z,{\bf q}_\|,\lambda}\big),
\end{equation}
{where $\hat{b}_{{\bf q},\lambda} = \frac{ R_{\bf{q}}}{| R_{\bf{q}}|} \hat{a}_{{\bf q},\lambda}$ and $\hat{b}_{{\bf q},\lambda}^\dagger = \frac{ R_{\bf{q}}^*}{| R_{\bf{q}}|} \hat{a}_{{\bf q},\lambda}^\dagger$.}

With the enhancement function defined, we choose the bins for the dispersion. However, we no longer choose to bin as a Cartesian grid in $q_x$ and $q_y$. Instead, we have a 3D dispersion with modes weighted by $R_{\bf q_z, \omega}$. The lossiness of the cavity causes the eigenmodes of the cavity to no longer be plane waves, which causes the broadening seen in Fourier plane absorption spectroscopy measurements. This can be interpreted as an uncertainty or broadening in $q_z = |{\bf q}_z |$ for a given frequency, $\omega$. Thus for a given mode order, $m$ and bin in $\omega$ and $\theta$, we group all modes with $q_z$ such that $m - 1/2 < q_z L_\mathrm{c} + \phi < m + 1/2$ in a bin, weighted by $R_{\bf q_z, \omega}$.

We then bin the vector potential under the assumption that $\omega\approx\omega_n, \forall \omega \in n$ in the bin and that ${\bf q}\approx {\bf q}_n, \forall {\bf q} \in n $ in the bin. The binned vector potential is then given by
\begin{align}
    \hat{\bf A}({\bf r}) =&~  \frac{1}{\sqrt{V}} \sum_{n,\lambda}  e^{i {\bf q}_{\|n} \cdot {\bf r}_\|} \hat{\boldsymbol{\epsilon}}_{{\bf q}_n,\lambda} A_{{\bf q}_n}  \nonumber \\ 
    &~~~~\times \frac{1}{\sqrt{2 \omega_n}} \sum_{\beta \in n}  R_\beta \,\, \big( \hat{b}_{\beta,\lambda}^\dagger +   \,\, \hat{b}_{-\beta,\lambda}\big),
\end{align}
where we define the superindex $\beta = \{ \bf{q}_\|, \omega\}$ such that $-\beta = \{ -\bf{q}_\|, \omega\}$ and also sums over each mode order.

As with the ideal case, we can then formulate the collective and relative coordinates for each bin as
\begin{subequations}
\begin{gather}
    \hat{X}_{n, \lambda} \equiv \sqrt{\frac{1}{2 \omega_n}} \sum_{\beta \in n} R_\beta \big( \hat{b}_{\beta,\lambda} + \hat{b}^{\dagger}_{-\beta,\lambda} \big)  = \sum_{\beta \in n} R_\beta \hat{x}_{\beta,\lambda} \\
    \hat{\chi}_{n,\beta,\lambda} \equiv \hat{x}_{\beta,\lambda} - \frac{R_\beta}{\ell_n} \hat{X}_{n, \lambda}  \\
    \hat{P}_{n, \lambda} \equiv - i \frac{\partial}{\partial \hat{X}_{n, \lambda}}\\
    \hat{\varphi}_{n,{\bf q},\lambda} \equiv \frac{\ell_n}{\ell_n-|R_{\bf q}|^2}\big( \hat{p}_{\beta, \lambda} -  R_\beta \hat{P}_{n, \lambda} \big), 
\end{gather}
\end{subequations}
where now each bin has its own $\ell_n$ as a weighted sum of all $R_\beta$ for $\beta \in n$,
\begin{align}
    \ell_n \equiv \sum_{\beta \in n} R_\beta^2.
\end{align}
Note that this is analogous to the number of modes in the bin for equally weighted modes, as in the ideal cavity case.

The effective photonic Hamiltonian then becomes,
\begin{align}
    \hat{H}_\mathrm{ph}^\mathrm{eff} =& \sum_{n,\lambda} \frac{1}{2} \big( \ell_n \hat{P}_{n, \lambda}^2 + \frac{1}{\ell_n} \omega_n^2 \hat{X}_{n, \lambda}^2 \big) + \hat{H}_\mathrm{ph}^\mathrm{rel} \\
    =& \sum_{n,\lambda} \hbar \omega_{n} \bigg( \hat{B}_{n,\lambda}^\dagger \hat{B}_{n, \lambda} + \frac{1}{2} \bigg) + \hat{H}_\mathrm{ph}^\mathrm{rel} \nonumber,
\end{align}
where $\hat{B}_{n,\lambda}^\dagger$ and $\hat{B}_{n, \lambda}$ are the creation and annihilation operators for the $n_\mathrm{th}$ bin's collective mode. 
This structure of collective mode operators is general for any binned vector potential, regardless of cavity geometry (see the SI).

{Similarly, the full Hamiltonian with respect to these effective modes takes an almost identical form to that of Eq.~\ref{eq:p.A-eff}}
\begin{align} \label{eq:p.A-eff_lossy}
    \hat{H}_\mathrm{p\cdot A} =& \hat{H}_\mathrm{el} + \sum_{n,\lambda} \hbar \omega_{n} \bigg( \hat{B}_{n,\lambda}^\dagger \hat{B}_{n, \lambda} + \frac{1}{2} \bigg) + \hat{H}_\mathrm{ph}^\mathrm{rel} \nonumber \\
    &+  \sum_{n,\lambda} \sqrt{\frac{\ell_n}{{V}}} \,\, A_{{\bf q}_n}  \hat{\bar{p}}_{n, \lambda} \big( \hat{B}_{n,\lambda} + \hat{B}^{\dagger}_{n,\lambda} \big) \nonumber \\
    &+  \sum_{n,\lambda,n',\lambda'} \frac{\sqrt{\ell_n \ell_{n'}}}{2V} \,\, \hat{\bar{s}}_{n, n'} A_{n} A_{n'} \hat{\boldsymbol{\epsilon}}_{n,\lambda} \cdot \hat{\boldsymbol{\epsilon}}_{n',\lambda'}   \nonumber \\
    &~~~~\times \big( \hat{B}_{n,\lambda} + \hat{B}^{\dagger}_{n,\lambda} \big) \big( \hat{B}_{n',\lambda'} + \hat{B}^{\dagger}_{n',\lambda'} \big), 
\end{align}
where $V / \ell_n$ is the effective mode volume of the $n_\mathrm{th}$ bin.

It is then useful to better understand $\ell_n / V$ and how the choice of bins affects the relative coupling strength of different effective modes, and verify that even for realistic cavities, the effective mode volume is independent of the mirror size. We can express $\ell_n / V$ as
\begin{align} \label{eq:ell_n}
    &\ell_n / V = \frac{1}{V} \sum_{{\bf q} \in n} | R_{{\bf q}}(z=0) |^2 \\
    &= \frac{1}{V} \sum_{{\bf q} \in n} \frac{1+2|\rflct| \cos(q_z L_\mathrm{c}+\phi)+|\rflct|^2}{1 + |\rflct|^4 - 2|\rflct|^2 \cos (2(q_zL_\mathrm{c}+\phi) )}  \nonumber\\
    &= \frac{dq_x \, dq_y \, dq_z}{8\pi^3 n_\mathrm{r}^3} \sum_{\{q_z,\omega, \theta\} \in n}   \frac{1+2|\rflct| \cos(q_z L_\mathrm{c}+\phi)+|\rflct|^2}{1 + |\rflct|^4 - 2|\rflct|^2 \cos (2(q_zL_\mathrm{c}+\phi) )} \nonumber\\
    &=\frac{1}{8\pi^3c^2 n_\mathrm{r}}\int_n dq_z \, d\omega \, d\theta \frac{\omega(1+2|\rflct| \cos(q_z L_\mathrm{c}+\phi)+|\rflct|^2)}{1 + |\rflct|^4 - 2|\rflct|^2 \cos (2(q_zL_\mathrm{c}+\phi) )}, \nonumber
\end{align}
{where $\phi$ is the complex phase of $\rflct$ which is usually close to $\pi$, and $\frac{n_\mathrm{r}^2\omega}{c^2}$ is the Jacobian transforming $(2 \pi)^3/V = dq_x \, dq_y \, dq_z$ to the coordinate basis of $\{ q_z, \omega, \theta \}$. We note that in order to match the continuous spectrum of modes in the free space limit ($\rflct \rightarrow 0$), the volume cannot be the volume of the cavity, and a limit of the volume going to infinity must be taken, justifying the integral limit of the sum.} {This integral can then be performed analytically\cite{Scherfgen_2026} to give}

\begin{align} \label{eq:ell_n_analytical}
    &\ell_n / V = \frac{[\theta]\big|_{\theta_{n,\mathrm{min}}}^{\theta_{n,\mathrm{max}}}[\omega^2]\big|_{\omega_{n,\mathrm{max}}}^{\omega_{n,\mathrm{max}}}}{16\pi^3c^2 n_\mathrm{r} L_\mathrm{c}\big||\rflct|-1\big|~\big||\rflct|+1\big|} \nonumber\\
    &\times\Bigg[\arctan\bigg(\frac{(|\rflct|^2+1)\tan(q_zL_\mathrm{c}+\phi)}{\big||\rflct|-1\big|~\big||\rflct|+1\big|}\bigg) \nonumber \\
    &+ \arctan\bigg(\frac{2|\rflct|\sin(q_zL_\mathrm{c}+\phi)}{\big||\rflct|-1\big|~\big||\rflct|+1\big|}\bigg)\Bigg]_{q_{z_\mathrm{min}}}^{q_{z_\mathrm{max}}}
\end{align}
{Defining $\Delta\omega_n=\omega_{n,\mathrm{max}}-\omega_{n,\mathrm{min}}$ and $\Delta\theta_n=\theta_{n,\mathrm{max}}-\theta_{n,\mathrm{min}}$ as the bin dimensions in $\omega$ and $\theta$ and deciding to bin $q_z$ over one full mode, and assuming $\phi\approx\pi$, Equation \ref{eq:ell_n_analytical} simplifies to give an effective volume}
\begin{equation}
    V_\mathrm{eff} =V/\ell_n \approx \frac{8\pi^2 c^2 n_\mathrm{r} L_\mathrm{c}\big||\rflct|-1\big|}{\omega_n \Delta\omega_n\Delta\theta}
\end{equation}
{for modes of odd $m$. 

For even modes, there is a node at the center of the cavity, and the integral is approximately equal to 0. In terms of quality factor, the effective volume is approximately}
\begin{equation}\label{eq:Veff}
    V^\mathrm{eff}_{n} \approx \frac{4\pi^3 c^2 n_\mathrm{r} L_\mathrm{c}m}{\omega_n \Delta\omega_n\Delta\theta_n {\mathcal Q}}.
\end{equation}
This suggests that the light matter coupling, which is proportional to $\sqrt{\ell_n/V}$, increases with better mirrors, which is intuitive but is in direct contrast to previous work\cite{Svendsen2025CP} that predicted a lower coupling strength with more reflective mirrors, where the crucial difference is that our theory accounts for the enhancement of modes rather than only counting the number of modes that are enhanced. Even though the cavity linewidth decreases with more reflective mirrors, causing fewer cavity modes to couple to the system, said modes couple more strongly to the matter such that the effective coupling strength of the bin increases rather than decreases with more reflective mirrors. It is also important to note that many of the parameters that show up are interrelated. Quality factor is inserted into the equation through its relation to $\rflct$. As such, the quality factor will only improve coupling if it is increased by changing the reflectivity of the mirrors. Increasing $\mathcal Q$ by any other means will not make improvements in coupling strength.

So far, the choice of $\Delta \omega_n$ and $\Delta\theta_n$ has been kept general. In the absence of coupling to matter, the plane-wave basis of photonic modes is orthogonal and perfectly defines every mode's frequency and momentum. The matter DOF, however, has some intrinsic uncertainty in both frequency and momentum. If we define $\hbar\Delta \omega_{\mathrm{mat},m}$ as the energetic uncertainty for the $m_\mathrm{th}$ matter state, then that state perceives two photons with frequencies $\omega_1 - \omega_0 < \Delta \omega_{\mathrm{mat},m}$ as having degenerate frequecies. As such, we can reasonably choose an upper bound of $\Delta \omega_n$ to be $\mathrm{min} (\Delta \omega_{\mathrm{mat},m})$ for all $m$ states that are energetically relevant to the $n_\mathrm{th}$ bin. Experimentally, we can approximate $\hbar\Delta \omega_{\mathrm{mat},m}$ as the linewidth of the $m_\mathrm{th}$ matter state's absorption peak (or equivalently via the Fourier transform of the state's lifetime).

Similarly, we propose to use the conservation of momentum between light and matter to inform the choice of $\Delta\theta_n$. Since the validity of creating these bins relies on the conservation of momentum as well as energy, the bins should all have approximately the same $\Delta \bf q$ to be under the same level of approximation. As such, we choose a different $\Delta \theta_n$ for each $\omega_n$ {to keep a constant $\Delta \bf q$}. We define $\Delta \theta_n$ as
\begin{equation}
    \Delta \theta_n =
    \begin{cases}
        \dfrac{2\pi}{\bigg\lfloor\dfrac{\pi}{\arcsin{ \left(\frac{\Delta q}{2q_\omega}\right) }}\bigg\rfloor}& \Delta q \le 2q_\omega\\
        2\pi&\Delta q > 2q_\omega
    \end{cases}
\end{equation}
{where $\Delta q$ is the uncertainty in momentum of the matter system due to its finite size and/or the presence of defects, and $\lfloor~ \rfloor$ denotes the floor function. The physical interpretation of this binning is that, with low uncertainty in momentum, the orientation of momentum is distinguishable, so the modes should be separated. The $\Delta \theta$ defined here is the orientation of momenta that are indistinguishable due to the Heisenberg uncertainty principle, with the uncertainty in position being set by the size of the crystal or of the molecules in the cavity. $\Delta \theta$ is the full $2\pi$ when the mode is perpendicular to the mirror, since ${\bf q} \ \cdot \hat{\bf z} = |{\bf q}|$. 

The attentive reader would note that from Eq.~\ref{eq:Veff} we get $V^\mathrm{eff}_{n} \propto \frac{1}{\Delta \omega \Delta \theta_n}$, which seems to imply that the light-matter interaction depends linearly on the choice of bin size. However, if the uncertainty of the matter frequency and momentum is such that $\Delta \omega_n < \Delta \omega_{\mathrm{mat}} $ and $\Delta \theta_n < \Delta \theta_{\mathrm{mat},n}$, shrinking the bin size further has no effect on the physical system. For example, if we halve the bin size, then we double the number of bins, $N$, and halve $V^\mathrm{eff}_{n}$. From the matter perspective, we can pair up all of these bins as pairs of indistinguishable frequency and momentum. As shown in the previous analysis, this is the same exact physical system as before the bins were halved. As such, for a perfectly modeled matter system, $\Delta \omega_n$ and $\Delta q$ can be treated as convergence parameters. However, for simulations that do not include matter loss channels or finite spatial descriptions, $\Delta \omega_n$ and $\Delta q$ should be phenomenologically set based on experimental results.



\begin{figure}[t]
    \centering
    \includegraphics[width=0.90\linewidth]{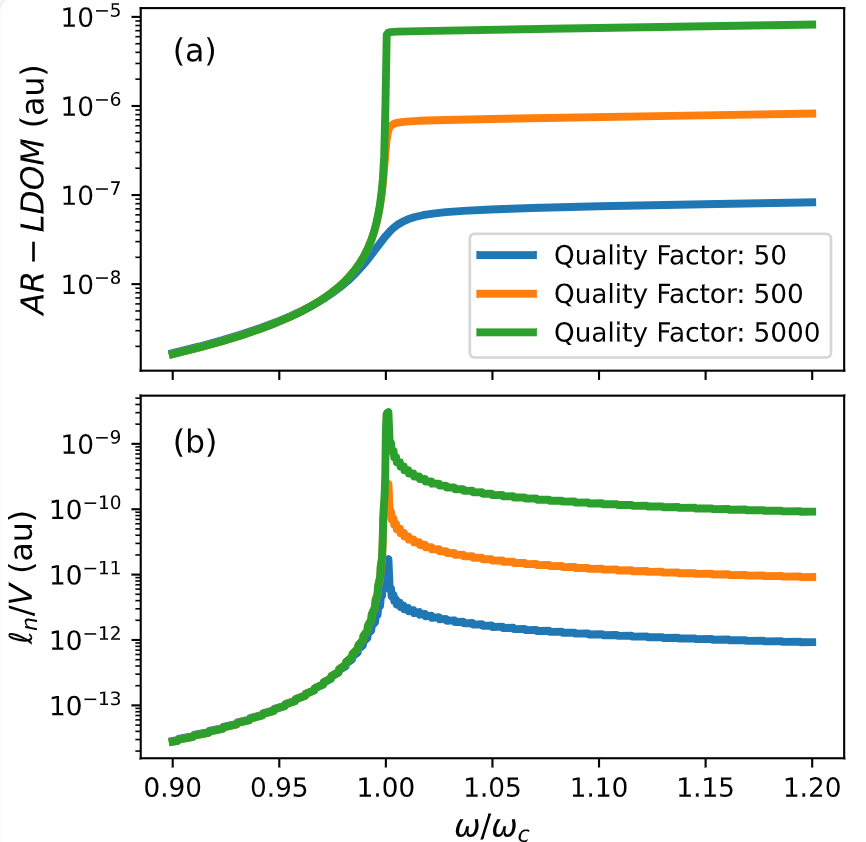}
    \caption{\small (a) AR-LDOM (See Eq.~\ref{eq:AR-LDOM}) and (b) $\ell_n$/$V$ with varying $\mathcal Q$ (See Eq.~\ref{eq:ell_n_analytical}). The fundamental mode of the cavity ($m = 1$)  is plotted with parameter values of $\omega_\mathrm{c} = 0.1$ au, $\Delta q=\frac{n_r\omega_c}{10c}$, and $\Delta \omega=\frac{\omega_c}{10}$, $n_\mathrm{r}=1$, and $\phi=\pi$.}
    \label{fig:ell_n_ARDOF}
\end{figure}

In the language of condensed matter physics, $\ell_n / V$, is reminiscent of a local density of states (LDOS); however, this terminology only works well in describing fermionic systems. For this bosonic system, we now define an \textit{ angular-resolved local density of modes} (AR-LDOM) as
\begin{align} \label{eq:AR-LDOM}
    D(\omega,\theta,z) &\equiv \frac{1}{8\pi^3 n_\mathrm{r}^3} \int_n d{\bf q} \, | R_{{\bf q}}(z) |^2 \delta (|{\bf q}| - \frac{n_\mathrm{r} \omega}{c} ) \nonumber\\ 
    &~~~~~~~~~~~~~~~~~\times\delta (\theta - \theta_{\bf q}) \nonumber \\
    &= \frac{\omega}{8\pi^3 n_\mathrm{r}^2 c} \int_n dq_z \, | R_{{\bf q}_z,\omega}(z) |^2,
\end{align}
{In general, an enhancement function would have $x,~y,~z, \mathrm{as ~well~as}~\theta$ dependence such that} the density of modes is not only spatially parameterized (local) but also angularly parameterized by the mode's in-plane direction (angular-resolved). It is apparent how this AR-LDOM is analogous to the LDOS from condensed matter physics, $D_\mathrm{s}(\omega,{\bf x})$, commonly defined as
    $D_\mathrm{s}(\omega,{\bf x}) \equiv \sum_j | \Psi_j ({\bf x})|^2  \delta (\omega - \omega_j)$, 
where $ \Psi_j ({\bf x})$ and $\omega_j$ are the wavefunction and energy of the $j_\mathrm{th}$ eigenstate, respectively. We can then write $\ell_n$ in terms of the AR-LDOM as
\begin{equation}
    \ell_n = \frac{n_\mathrm{r}V}{c} \iint_n d\omega \, d\theta \, D(\omega, \theta, 0).
\end{equation}
In the limit of $\{ \Delta \omega_n, \Delta\theta_n\} \to 0$, $g(\omega_n, \theta_n,z) \equiv  \ell_n / V$ is the angular-resolved local degree of degeneracy. In other words, we show that the light-matter coupling strength scales {as $\sqrt{g(\omega_n, \theta)}$.

Fig.~\ref{fig:ell_n_ARDOF} plots both the AR-LDOM and $\frac{\ell_n}{V}$ using this method for defining $\Delta \theta$. As a direct consequence of ensuring that all bins have the same uncertainty in momentum, $\frac{\ell_n}{V}$ has a sharp peak at $\omega = \omega_\mathrm{c}$ whose magnitude is approximately an order of magnitude greater than that of the high-frequency limit. This implies that for systems highly sensitive to momentum, unique phenomena might occur as $|{\bf q}|_\| \to 0$.
It should also be noted here that although the binning in $\theta$ may create a small value of $\frac{\ell_n}{V}$ at frequencies higher than $\omega_c$, there is a proportionate number of modes that are equivalent in all ways except orientation of momentum. Many systems have larger uncertainty than the momentum of the photons being absorbed, so in these cases, $\Delta\theta$ should be the full $2\pi$, and $\ell_n/V$ would just be proportional to the AR-LDOM with no peak at resonance. Investigating how this momentum-uncertainty-dependent resonance effect might occur in systems like low-defect crystals and Bose-Einstein condensates remains an exciting avenue for further studies.}

\section{\large Conclusions} \label{swec:concl}
In this paper, we introduced the effective modes framework for a better understanding of the photonic degrees of freedom for cavity QED in the strong coupling regime. This new methodology can be applied ubiquitously across all of theoretical cavity QED, as it is ambivalent to the gauge choice or matter DOF.

In this work, we first set the stage by discussing the minimal coupling Hamiltonian under second-quantization. By writing the Hamiltonian in this form, we explicitly see the conservation of momentum terms between the light and matter DOF. Additionally, we discuss how exact physical models must include a quasi-continuum of photonic modes, which inherently makes directly diagonalizing the Hamiltonian computationally intractable. These properties of the exact Hamiltonian provide the background and motivation for the effective modes framework.

Then, we introduce the effective modes framework on the simplest non-trivial model cavity, an ideal Fabry-P\'erot cavity. Instead of coarsely sampling photonic modes from the quasi-continuous dispersion as commonly done in the literature, we bin together modes of similar $\bf q$ and $\omega$ to form effective modes. We then perform a change of basis for each bin to reduce the many modes coupled to the matter to a single effective mode coupled to the matter. This enables us to define a quantization volume for these effective modes, which is completely independent of the mirror size. While a convenient, simple model, this ideal cavity relies on an ad-hoc choice of bin size, so a more realistic cavity needs to be considered.

Finally, we expand our formalism to such a realistic cavity, where we explicitly consider a Fabry-P\'erot cavity with lossy mirrors. In such a case, there are no exactly forbidden modes, so the dispersion relation becomes $\omega = c |{\bf q}| / n_\mathrm{r}$; however, the cavity enhances or retards the field strength of these modes. As such, we refine the effective modes framework to integrate over this spatially varying field enhancement factor $R_{\bf q} (z)$, enabling us to rigorously understand how the mirror reflectivity affects the light-matter coupling through the quantization volume of the effective modes. This formalism also resembles the local density of states from condensed matter physics, but for this bosonic subsystem, we also parameterize by the azimuthal angle of $\bf q$, creating the angular-resolved local density of modes (AR-LDOM).



This work has the potential to enable many future works in the field of cavity QED as a whole. The ideal cavity case allows us to better understand the implicit approximations made when considering only a few photonic modes along a dispersion, and can be directly applied to any lossless model that is sampling the dispersion relation. Additionally, while we only consider Fabry-P\'erot cavities, this formalism can be easily applied to more exotic cavity geometries such as plasmonic cavities, since it is general for any form of $R_{\bf q}$. This work also provides a new understanding of mode volume.
}

\section*{\large Acknowledgement}
{\footnotesize This material is based upon work supported by the W.M. Keck Foundation and by the Air Force Office of Scientific Research under AFOSR Award No. FA9550-23-1-0438. M.T. appreciates the support from the National Science Foundation Graduate Research Fellowship Program under Grant No. DGE-1939268. Computing resources were provided by the Center for Integrated Research Computing (CIRC) at the University of Rochester. 

{\bf M.A.D.T. and J.S. contributed equally to this work.}}

{\footnotesize
\bibliography{references.bib}
}

\end{document}


\setcounter{section}{0}
\setcounter{subsection}{0}
\setcounter{subsubsection}{0}
\setcounter{equation}{0}
\setcounter{figure}{0}

\renewcommand{\theequation}{S\arabic{equation}}
\renewcommand{\thesection}{S\arabic{section}}
\renewcommand\thefigure{S\arabic{figure}}

\section{\normalsize Derivation of Effective Modes with spatial dependence} \label{app:R_q}

Vector potential enhanced by a cavity with an enhancement function $R_{\bf q}({\bf r})$ (where we have explicitly allowed for the enhancement function to have spatial dependence) is given by

\begin{equation}
    \hat{\bf A}({\bf r}) = \sum_{{\bf q},\lambda}   \hat{\boldsymbol{\epsilon}}_{{\bf q},\lambda} \sqrt{\frac{\hbar}{ 2\varepsilon_{0} \omega_{\bf q} {V}}}   \big( R_{\bf q}({\bf r}) \hat{a}_{{\bf q},\lambda}e^{i {\bf q}\cdot {\bf r}} + \mathrm{h.c.} \big).
\end{equation}

{We rotate out the phase in the enhancement function by defining $\hat{b}_{{\bf q},\lambda} \equiv \frac{R_{\bf q}({\bf r})}{|R_{\bf q}({\bf r})|} \hat{a}_{{\bf q},\lambda}$. We then bin the vector potential under the assumption that $\omega\approx\omega_n$ for all $\omega$ in the bin (photon energy is approximately constant) and that ${\bf q}\approx {\bf q}_n$ for all ${\bf q}$ in the bin (polarization and spatial phase of the vector potential are approximately constant). We note that the enhancement function may change sharply with small changes in wavevector, so we don't assume that $R_{\bf q}({\bf r})$ is constant over a bin. The binned vector potential is then given by
\begin{equation}
    \hat{\bf A}({\bf r}) \approx \sum_{n,\lambda} \sum_{{\bf q}\in n}  \hat{\boldsymbol{\epsilon}}_{{\bf q}_n,\lambda} \sqrt{\frac{\hbar}{ 2\varepsilon_{0} \omega_{n} {V}}}   \big( |R_{\bf q}({\bf r})| \hat{b}_{{\bf q}_n,\lambda}e^{i {\bf q}_n\cdot {\bf r}} + \mathrm{h.c.} \big).
\end{equation}

Instead of reorganizing terms as was done in the paper to factor the spatial dependence out, we instead define a collective coordinate with explicit spatial dependence as

\begin{equation}
    \hat{X}_{n, \lambda}({\bf r}) \equiv \sqrt{\frac{\hbar}{2 \omega_n}} \sum_{{\bf q} \in n} |R_{\bf q}({\bf r})| \big( e^{-i\bf q_n\cdot r} \hat{b}_{{\bf q},\lambda}^\dagger + \mathrm{h.c.} \big) = \sum_{{\bf q} \in n} |R_{\bf q}({\bf r})| \hat{x}_{{\bf q},\lambda}({\bf r}) 
\end{equation}

We note that this coordinate is not normalized and has an infinitely large effect. In the paper, the $p\cdot A$ Hamiltonian only depends upon the creation and annihilation operators, and normalization was done when effective mode operators were formed. More generally, here, we normalize the collective coordinate and momentum by letting
\begin{equation}
    \hat{x}_{n, \lambda}({\bf r}) \equiv \frac{1}{\sqrt{\ell_n({\bf r})}} \hat{X}_{n, \lambda}({\bf r}),
\end{equation}
where $\ell_n({\bf r})$ is not yet derived but will be shown to be equivalent to the $\ell_n$ given in the paper.

Since $\hat{x}_{{\bf q},\lambda}({\bf r})$ can be thought of as directions in a hyperdimensional space, we can switch each $\hat{x}_{{\bf q},\lambda}({\bf r})$ and the normalized collective coordinate, $\hat{x}_{n, \lambda}({\bf r})$, to a corresponding unit vector in its direction. The dot product between two unit vectors is 1, so the unit vector corresponding to $\hat{x}_{n, \lambda}({\bf r})$ can be used to define the normalization constant. With the same notation as operators but now thinking of each $\hat{x}$ as a unit vector and realizing that each $\hat{x}_{{\bf q},\lambda}({\bf r})$ is an orthogonal direction, we get
\begin{align}
    \hat{x}_{n, \lambda}({\bf r})\cdot \hat{x}_{n, \lambda}({\bf r}) &= \frac{1}{\ell_n({\bf r})} \sum_{{\bf q}, {\bf q'} \in n} |R_{\bf q}({\bf r})| |R_{\bf q'}({\bf r})| \hat{x}_{{\bf q},\lambda}({\bf r}) \cdot \hat{x}_{{\bf q'},\lambda}({\bf r}) \nonumber \\
    1 &= \frac{1}{\ell_n({\bf r})} \sum_{{\bf q}, {\bf q'} \in n} |R_{\bf q}({\bf r})| |R_{\bf q'}({\bf r})| \delta_{\bf q,q'} \nonumber \\
    1 &= \frac{1}{\ell_n({\bf r})} \sum_{{\bf q} \in n} |R_{\bf q}({\bf r})|^2 \nonumber \\
    \ell_n({\bf r}) &= \sum_{{\bf q} \in n} |R_{\bf q}({\bf r})|^2
\end{align}

Switching back to thinking about operators, and with the normalized collective coordinate now fully defined, we can write the collective and relative coordinates and momenta as
\begin{subequations}
\begin{gather}
    \hat{x}_{n, \lambda}({\bf r}) = \frac{1}{\sqrt{\ell_n({\bf r})}} \sum_{{\bf q} \in n} |R_{\bf q}({\bf r})| \hat{x}_{{\bf q},\lambda}({\bf r}) \nonumber\\
    \hat{p}_{n, \lambda}({\bf r}) \equiv -i\hbar\frac{\partial}{\partial \hat{x}_{n, \lambda}({\bf r})} =  \frac{1}{\sqrt{\ell_n({\bf r})}} \sum_{q \in n} |R_{\bf q}({\bf r})| \hat{p}_{{\bf q},\lambda}({\bf r})\\
    \hat{\chi}_{n,{\bf q},\lambda}({\bf r}) \equiv \sqrt{\frac{\ell_n({\bf r})}{\ell_n({\bf r})-|R_{\bf q}({\bf r})|^2}} \bigg(\hat{x}_{{\bf q},\lambda}({\bf r}) - \frac{|R_{\bf q}({\bf r})|}{\sqrt{\ell_n({\bf r})}} \hat{x}_{n, \lambda}({\bf r})\bigg) \\
    \hat{\varphi}_{n,{\bf q},\lambda}({\bf r}) \equiv \sqrt{\frac{\ell_n({\bf r})}{\ell_n({\bf r})-|R_{\bf q}({\bf r})|^2}} \bigg(\hat{p}_{{\bf q}, \lambda}({\bf r}) -  \frac{|R_{\bf q}({\bf r})|}{\sqrt{\ell_n({\bf r})}} \hat{p}_{n, \lambda}({\bf r})\bigg).
\end{gather}
\end{subequations}}
The photonic Hamiltonian is written as
\begin{align}
    \hat{H}_\mathrm{ph} = \sum_{\lambda,{\bf q}} \hbar\omega_{\bf q}\bigg(\hat{a}_{{\bf q},\lambda}^\dagger \hat{a}_{{\bf q},\lambda}+\frac{1}{2}\bigg)
    = \sum_{\lambda,{\bf q}} \hbar \omega_{{\bf q}} \bigg(\hat{b}_{{\bf q},\lambda}^\dagger \hat{b}_{{\bf q},\lambda}+\frac{1}{2}\bigg)
    \approx \sum_{n,\lambda,{\bf q} \in n} \hbar \omega_{{\bf q}_n} \bigg(\hat{b}_{{\bf q},\lambda}^\dagger \hat{b}_{{\bf q},\lambda}+\frac{1}{2}\bigg) 
\end{align}
The spatial dependence in $\hat{x}_{{\bf q},\lambda}({\bf r})$ and $\hat{p}_{{\bf q},\lambda}({\bf r})$ cancels out when squared and combined such that the photonic Hamiltonian can be written as
\begin{align}\label{eq:Hpheff}
    \hat{H}_\mathrm{ph} =& \sum_{n,\lambda,{\bf q} \in n} \hbar \omega_{{\bf q}_n} \bigg(\hat{b}_{{\bf q},\lambda}^\dagger \hat{b}_{{\bf q},\lambda}+\frac{1}{2}\bigg)
    =\sum_{n,\lambda,{\bf q} \in n} \frac{1}{2} (\hat{p}_{{\bf q}, \lambda}^2({\bf r}) + \omega_n^2 \hat{x}_{{\bf q}, \lambda}^2({\bf r}))\nonumber\\
    =&\sum_{n,\lambda,{\bf q} \in n} \frac{1}{2}\left(\frac{|R_{\bf q}({\bf r})|^2}{\ell_n({\bf r})}\right) \Big( \hat{p}_{n, \lambda}^2({\bf r}) + \omega_n^2 \hat{x}_{n, \lambda}^2({\bf r}) \Big) 
    + \sum_{n,\lambda,{\bf q} \in n} \frac{1}{2} \bigg(\frac{\ell_n({\bf r})-|R_q({\bf r})|^2}{\ell_n({\bf r})}\bigg) \Big(\hat{\varphi}_{n,{\bf q},\lambda}^2({\bf r}) + \omega_n^2 \hat{\chi}_{n,{\bf q},\lambda}^2({\bf r}) \Big) \nonumber \\
    =& \sum_{n,\lambda} \frac{1}{2} \Big( \hat{p}_{n, \lambda}^2({\bf r}) + \omega_n^2 \hat{x}_{n, \lambda}^2({\bf r}) \Big) + \hat{H}_{ph}^{rel},
\end{align}
where the cross terms, containing $\hat{\chi}_{n,{\bf q},\lambda}({\bf r}) \hat{x}_{n, \lambda}({\bf r})$ and $\hat{\varphi}_{n,{\bf q},\lambda}({\bf r}) \hat{p}_{n, \lambda}({\bf r})$, equal zero due to the strategic choice of coefficients in $\hat{\chi}_{n,{\bf q},\lambda}({\bf r})$ and $\hat{\varphi}_{n,{\bf q},\lambda}({\bf r})$ that makes $\sum_{{\bf q} \in n} |R_{\bf q}({\bf r})| \hat{\chi}_{n,{\bf q},\lambda}({\bf r}) = 0$ and $\sum_{{\bf q} \in n} |R_{\bf q}({\bf r})| \hat{\varphi}_{n,{\bf q},\lambda}({\bf r}) = 0$.

We now form effective creation and annihilation operators for each bin by letting
\begin{equation}
    \hat{x}_{n, \lambda}({\bf r}) = \sqrt{\frac{\hbar}{2 \omega_n}} \big( e^{-i\bf q_n\cdot r} \hat{B}_{n,\lambda}^\dagger + h.c. \big).
\end{equation}
Comparing definitions of $\hat{x}_{n, \lambda}({\bf r})$ results in the effective creation and annihilation operators being defined in terms of individual ones as $\hat{B}_{n,\lambda} = \frac{1}{\sqrt{\ell_n({\bf r})}} \sum_{q \in n} |R_{\bf q}({\bf r})| \hat{b}_{q,\lambda}$. This definition is equivalent to forming effective operators by normalizing according to a weighted average. Plugging this definition for $\hat{x}_{n, \lambda}({\bf r})$ into Equation \ref{eq:Hpheff} gives
\begin{equation}
    \hat{H}_\mathrm{ph} = \sum_{n,\lambda} \hbar \omega_{n} \bigg( \hat{B}_{n,\lambda}^\dagger \hat{B}_{n, \lambda} + \frac{1}{2} \bigg) + \hat{H}_\mathrm{ph}^\mathrm{rel},
\end{equation}
The vector potential can now also be written in terms of these effective operators as
\begin{equation}
    \hat{\bf A}({\bf r}) \approx \sum_{{n},\lambda}   \hat{\boldsymbol{\epsilon}}_{{n},\lambda} \sqrt{\frac{\hbar \ell_n({\bf r})}{ 2\varepsilon_{0} \omega_{n} {V}}}   \big(\hat{B}_{{n},\lambda}e^{i {\bf q}_n\cdot {\bf r}} + h.c \big)
\end{equation}
This looks identical to the textbook definition of the vector potential, but now with the volume becoming an effective volume given by $V_\mathrm{eff}=\frac{V}{\ell_n({\bf r})}$. Note that, other than the assumption that the wavevectors and frequencies in each bin are approximately constant, this derivation has been kept general. As such, once the enhancement function has been derived for a system, any system's Hamiltonian and vector potential can be approximated in this way (see Supporting Information for a derivation of the d$\cdot$E Hamiltonian).

\section{\normalsize Deriving $R_{\bf q} (z)$ for a Fabry-P\'{e}rot Cavity} \label{app:R_q}

Suppose we have an asymmetric Fabry-P\'{e}rot cavity with flat, aligned mirrors having reflectivities of $\rflct_1$ and $\rflct_2$ separated by a distance $L_\mathrm{c}$ along the z-direction with the origin placed at the center of the cavity. An incident plane wave of the form $\frac{1}{\sqrt{2\pi}} e^{i {\bf q}_r \cdot \bf r}e^{-\frac{q_{i}}{\cos(\alpha)}(\frac{L_\mathrm{c}}{2}+z)}$ becomes $\frac{1}{\sqrt{2\pi}} \rflct e^{-i {\bf q}_{zr}\cdot \bf z} e^{i {\bf q_{\|r}}\cdot \bf r_\|} e^{-\frac{q_{i}}{\cos(\alpha)}(\frac{L_\mathrm{c}}{2}-z)}$ upon reflection (where the angular wavenumber is defined as $q\equiv q_{r}+iq_{i}$ and $\alpha$ is the angle of incidence with $\cos(\alpha)= \frac{q_{zr}}{q_r}$). In order to understand the enhancement factor, $R_{\bf q} (z)$, we start with a particle located at $z_0$ emitting a photon into an eigenmode of the vacuum with a quantization box volume of $V$
\begin{equation}
    {\bf U}_{{\bf q}, \lambda,0} = \frac{A_{\bf q}}{\sqrt{(2 \pi)^3 V}} \hat{\boldsymbol{\epsilon}}_{{\bf q}, \lambda} e^{i \bf q_r \cdot r} e^{-\frac{q_{i}}{\cos(\alpha)}(-z_0+z)}.
\end{equation}
Assuming homogeneous materials, upon a single reflection with a cavity mirror normal to the z-direction, this plane wave becomes
\begin{align}
    {\bf U}_{{\bf q}, \lambda,0}' =& \frac{\rflct_1 A_{\bf q}}{\sqrt{(2 \pi)^3 V}} \hat{\boldsymbol{\epsilon}}_{-q_z, {\bf q_\|}, \lambda} e^{i ({\bf q_{\|r} \cdot r_\|}-{\bf q}_{zr} \cdot {\bf z} + |q_{zr}|L_\mathrm{c})} \nonumber \\
    &\times e^{-  \frac{q_{i}}{\cos(\alpha)} (\frac{L_\mathrm{c}}{2} - z) } e^{-  \frac{q_{i}}{\cos(\alpha)} (\frac{L_\mathrm{c}}{2}-z_0)}.
\end{align}
The absorption factor $e^{-\frac{q_{i}}{\cos(\alpha)} (\frac{L_\mathrm{c}}{2}-z_0)}$ from this first reflection is the only one of its kind. All subsequent reflections with the mirrors pick up an absorption factor of $e^{-\frac{q_{i}}{\cos(\alpha)} L_\mathrm{c}}$. If there will be many reflections, $q_i$ is small, or $z_0\approx -\frac{L_\mathrm{c}}{2}$, the math can be simplified with close to the same end result by letting $e^{-\frac{q_{i}}{\cos(\alpha)} (\frac{L_\mathrm{c}}{2}-z_0)} \approx e^{-\frac{q_{i}}{\cos(\alpha)} L_\mathrm{c}}$. Under this approximation, after a single round trip in the cavity, the field becomes,
\begin{equation}
    {\bf U}_{{\bf q}, \lambda,1} = \rflct_1 \rflct_2 e^{2 i |q_{zr}| L_c } e^{-2 q_{zi} L_c} {\bf U}_{{\bf q}, \lambda,0}.
\end{equation}
By adding together the amplitudes from a finite number of double reflections ($N_d$), we get
\begin{align}
&{\bf U}_{{\bf q}, \lambda} = {\bf U}_{{\bf q}, \lambda,0} + {\bf U}_{{\bf q}, \lambda,0}' +  {\bf U}_{{\bf q}, \lambda,1} +  {\bf U}_{{\bf q}, \lambda,1}' + \, \cdots + {\bf U}_{{\bf q}, \lambda,N_d} + {\bf U}_{{\bf q}, \lambda,N_d}'\nonumber\\
&= {\bf U}_{{\bf q}, \lambda,0} + {\bf U}_{{\bf q}, \lambda,0}' + \rflct_1 \rflct_2 e^{2 i |q_{zr}| L_c} e^{-2 q_{zi} L_c} ({\bf U}_{{\bf q}, \lambda,0} + {\bf U}_{{\bf q}, \lambda,0}') \nonumber \\
&~~~~~~~~~~~~+ \cdots + (\rflct_1 \rflct_2)^{N_d} e^{2N_d i |q_{zr}| L_c} e^{-2 N_d q_{zi} L_c} ({\bf U}_{{\bf q}, \lambda,0} + {\bf U}_{{\bf q}, \lambda,0}') \nonumber \\
&= \frac{({\bf U}_{{\bf q}, \lambda,0} + {\bf U}_{{\bf q}, \lambda,0}')(1 - (\rflct_1 \rflct_2)^{(N_d+1)} e^{2(N_d+1) i |q_{zr}| L_c} e^{-2(N_d+1) q_{zi} L_c})}{1 - \rflct_1 \rflct_2 e^{2 i |q_{zr}| L_c} e^{-2 q_{zi} L_c}} \nonumber \\
&=  \frac{A_{\bf q} e^{i \bf q_{\|r} \cdot r_\|}}{\sqrt{(2 \pi)^3 V}} \bigg(\hat{\boldsymbol{\epsilon}}_{{\bf q}, \lambda} e^{i \bf q_{zr }\cdot z} e^{-\frac{q_{i}}{\cos(\alpha)}(-z_0+z)} + \rflct_1 \hat{\boldsymbol{\epsilon}}_{-q_z, {\bf q_\|}, \lambda} e^{-i {\bf q}_{zr} \cdot {\bf z}} e^{i |q_{zr}|L_\mathrm{c}} e^{-  \frac{q_{i}}{\cos(\alpha)} (\frac{3L_\mathrm{c}}{2} - z) }\bigg) \,\,\nonumber \\
&~~\times\bigg(\frac{1 - (\rflct_1 \rflct_2)^{(N_d+1)} e^{2(N_d+1) i |q_{zr}| L_c} e^{-2(N_d+1) q_{zi} L_c}}{1 - \rflct_1 \rflct_2 e^{2 i |q_{zr}| L_c} e^{-2 q_{zi} L_c}}\bigg).
\end{align}

This assumes there are exactly an even number of reflections. With many reflections, the error due to this approximation would be small, but if more exact results are needed, a term could be excluded from the geometric series or one could explicitly do the full sum.

As long as all materials are isotropic such that the refractive indices do not need to be defined as second rank tensors, TE and TM polarizations will not change polarization upon reflection. Thus, when decomposed into its TE and TM components, $\hat{\boldsymbol{\epsilon}}_{{\bf q}, \lambda} = \hat{\boldsymbol{\epsilon}}_{-q_z, {\bf q_\|}, \lambda}$, and thus $\bar{\bf U}_{{\bf q}, \lambda}$ and ${\bf U}_{{\bf q}, \lambda,0}$ have the same polarization. With this, and setting $z=z_0$, we can define the enhancement function $R_{\bf q}$ as 
\begin{align}\label{eq:asymmetric1}
    R_{\bf q} &= \frac{\bar{U}_{{\bf q}, \lambda}}{{ U}_{{\bf q}, \lambda,0}} = \bigg( 1 + \rflct_1  e^{-2i ({\bf q}_{zr} \cdot {\bf \hat{z}})z_0} e^{i |q_{zr}|L_\mathrm{c}} e^{-  \frac{q_{i}}{\cos(\alpha)} (\frac{3L_\mathrm{c}}{2} -z_0) }\bigg) \,\,\nonumber \\
&~~\times\bigg(\frac{1 - (\rflct_1 \rflct_2)^{(N_d+1)} e^{2(N_d+1) i |q_{zr}| L_c} e^{-2(N_d+1) q_{zi} L_c}}{1 - \rflct_1 \rflct_2 e^{2 i |q_{zr}| L_c} e^{-2 q_{zi} L_c}}\bigg).
\end{align}
For a photon emitted in the opposite z-direction
\begin{align}\label{eq:asymmetric2}
    R_{\bf q} &= \bigg( 1 + \rflct_2  e^{-2i ({\bf q}_{zr} \cdot {\bf \hat{z}})z_0} e^{i |q_{zr}|L_\mathrm{c}} e^{-  \frac{q_{i}}{\cos(\alpha)} (\frac{3L_\mathrm{c}}{2} + z_0) }\bigg) \,\,\nonumber \\
&~~\times\bigg(\frac{1 - (\rflct_1 \rflct_2)^{(N_d+1)} e^{2(N_d+1) i |q_{zr}| L_c} e^{-2(N_d+1) q_{zi} L_c}}{1 - \rflct_1 \rflct_2 e^{2 i |q_{zr}| L_c} e^{-2 q_{zi} L_c}}\bigg).
\end{align}

The value $N_d$ is a function of geometry and particle location. For circular mirrors and neglecting effects of diffraction, we get
\begin{equation}
    N_d=\Bigg\lfloor\frac{l |q_z|}{2L_\mathrm{c}\sqrt{\frac{\omega^2}{c^2} - q_z^2}}-\frac{({\bf q_z \cdot \hat{z}})z_0}{2 |q_z| L_\mathrm{c}} + 0.25\Bigg\rfloor
\end{equation}
where $\lfloor~\rfloor$ denotes the floor function, and
\begin{equation}
    l=\sqrt{r_\mathrm{c}^2+x_0^2 + y_0^2 - 2r_\mathrm{c}(x_0\cos(\theta) + y_0\sin(\theta))}
\end{equation}
where $r_\mathrm{c}$ is the radius of the mirrors, and the particle is located at $(x_0,y_0,z_0)$.

In the main body of the paper, we assume a symmetric mirror setup such that $\rflct_1=\rflct_2=\rflct$. We also assume the mirrors are large enough that there will be many reflections ($N_d\rightarrow\infty$). The final approximation made is that $q_i\approx0$. With these approximations, the enhancement function goes to what is given in the paper
\begin{equation}
    R_{\bf q} = \frac{1 + \rflct e^{i (|q_{zr}|L_\mathrm{c} - 2 {\bf q}_{zr} \cdot {\bf z})} }{1 - \rflct^2 e^{2 i |q_{zr}| L_c} }.
\end{equation}
We note that these enhanced modes are not strictly orthogonal, but they do span the space and can be reasonably approximated to be orthogonal since the modes are fully orthogonal outside the cavity where a majority of the field lies.

\section{\normalsize Vacuum and Ideal Cavity Limits} \label{app:limits}

This theory should be valid in the vacuum limit ($\rflct \rightarrow 0$) and the ideal cavity limit ($\rflct \rightarrow -1$ and $r_\mathrm{c} \rightarrow \infty$). The vacuum limit is easy to see. If $\rflct_1\rightarrow0$ and $\rflct_2\rightarrow0$ in Eq. \ref{eq:asymmetric1}, we get that
\begin{equation}
    R_{\bf q} = 1,
\end{equation}
which is exactly the vacuum where nothing is enhanced or suppressed, and the vector potential becomes the typical free-space vector potential after field quantization.

The ideal cavity limit is trickier. To simplify the analysis, we go to a symmetric system such that $\rflct_1=\rflct_2=|\rflct|e^{i\phi}$ and let absorption within the cavity be small such that $q_\mathrm{zi}\rightarrow0$. Under these conditions and taking the limit $r_\mathrm{c} \rightarrow \infty$, we get
\begin{align}\label{eq:FullSymmetricRsqr}
    |R_{\bf q}|^2 &= \frac{1 + |\rflct|^2+2|\rflct|  \cos\big(q_{z}L_\mathrm{c}+\phi-2 q_\mathrm{z}z\big)}{1 +|\rflct|^4 - 2|\rflct|^2 \cos\big(2 (q_\mathrm{z} L_c + \phi)\big)}.
\end{align}
The ideal cavity as used in the paper would require that the continuous limit be taken and that $|R_{\bf q}|^2 = \delta({q}_\mathrm{z}L_\mathrm{c} - m\pi)$. It is clear that $|R_{\bf q}|^2$ does approach infinity at ${q}_\mathrm{z}L_\mathrm{c}= m\pi$ in the limit $\rflct\rightarrow-1$. The function would then need to be normalized to make the result of the integral 1 for the limit to be reached. We normalize our functions as
\begin{equation}
    \int_V \epsilon({\bf r})|{ R}_{\bf q}({\bf r})|^2 d^3r = \epsilon_0V
\end{equation}
where $V$ is the entire volume inside and outside the cavity and $R_q$ is assumed to be roughly 1 outside the cavity. For this equation to be true, we need to divide the definition of $R_{\bf q}$ given in Eq. \ref{eq:FullSymmetricRsqr} by the factor
\begin{align}
    \frac{\int_{V_\mathrm{cav}} \epsilon({\bf r})|{ R}_{\bf q}({\bf r})|^2 d^3r +\int_{V-V_\mathrm{cav}} \epsilon_0 d^3r }{ \epsilon_0V} = 1+\frac{\int_{V_\mathrm{cav}} \epsilon({\bf r})|{ R}_{\bf q}({\bf r})|^2 d^3r -V_\mathrm{cav} \epsilon_0 }{ \epsilon_0V}
\end{align}
In the limit as $V\rightarrow\infty$, the second term is approximately zero, and we get
\begin{equation}
    1+\frac{\int_{V_\mathrm{cav}} \epsilon({\bf r})|{ R}_{\bf q}({\bf r})|^2 d^3r -V_\mathrm{cav} \epsilon_0 }{ \epsilon_0V}\approx1.
\end{equation}
Thus, in the infinite volume limit, there is no need to renormalize the wave function even though the field is enhanced within the cavity. For large but finite volumes of free space, the field would be renormalized to account for larger enhancements, and the ideal cavity limit is well-defined, but the point where renormalization is non-negligible would be highly dependent on the exact size of free space. Thus, we assume free space is large enough that no renormalization is necessary. From this perspective, there is a discontinuity between the lossy and ideal cavity because only at an ideal cavity, where the enhancement is infinite and can compete with the infinite volume, would you need to renormalize. This can also be thought of as there being a fundamental difference between a lossy cavity which is an open system connected to free space, and an ideal cavity which is a fully closed system.

There has been some discussion in literature on the correct form of mode volume. A common approximate form of mode volume is given by \cite{Coccioli1998IEE}
\begin{equation}
    V=\frac{\int\epsilon({\bf r})E^2({\bf r})d^3{\bf r}}{\mathrm{max}[\epsilon({\bf r})E^2({\bf r})]}.
\end{equation}
Leung et al. \cite{Ho1998PRE,Leung1994PRA} give a much more accurate form of mode volume and point out that a proper normalization of modes needs to include surface waves and take the limit as $V\rightarrow\infty$ to account for the full electric field and leakage outside the cavity. Muljarov et al. \cite{Muljarov2016PRB} improve upon this normalization definition but find that the limit as $V\rightarrow\infty$ diverges, so they normalize by integrating only over the inhomogeneous region of space. 
Our analysis takes a different view by saying that the cavity is so much smaller than the free-space volume $V$ such that the effect of the cavity enhancement on the normalization condition can be ignored and no further renormalization of the modes is necessary.
It is this difference in normalization that changes the definition of mode volume from being related to physical volume to being a measure of coupling strength as found in this paper.



\section{\normalsize Deriving the d$\cdot$E and Pauli Fierz Hamiltonians} \label{app:ddotE}

The p$\cdot$A Hamiltonian is given by

\begin{align}\label{eq:p.A}
    \hat{H}_\mathrm{p\cdot A} =&~ \sum_i^{N_e} \frac{1}{2m} \Big( \hat{\bf p}_i - e\hat{\bf A}(\hat{\bf r}_i) \Big)^2 + \frac{1}{2} \sum_{i \neq j}^{N_e} v_\mathrm{ee}(\hat{r}_i,\hat{r}_j) + \sum_i^{N_e} v_\mathrm{eN}(\hat{r}_i) + \sum_{\bf{q},\lambda} \hbar \omega_{\bf{q}} \Big(\hat{a}_{\bf{q},\lambda}^\dagger \hat{a}_{\bf{q},\lambda} + \frac{1}{2} \Big).
\end{align}

To transform to d$\cdot$E, we use the unitary transformation defined by

\begin{equation}
    U=e^{-\frac{i}{\hbar} \sum_{j=1}^N {\bf \vec{\bf \mu}_j\cdot\vec{A}(\overline{{\bf r}}_j)}},
\end{equation}
where $\vec{\bf \mu}_j$ is the dipole moment and $\overline{{\bf r}}_j$ is the center of mass of molecule $j$. With this unitary defined, the p$\cdot$A Hamiltonian can also be written as\cite{Mandal2023CR}

\begin{equation}
    \hat{H}_\mathrm{p\cdot A}=\hat{U}^\dagger\hat{H}_\mathrm{M}\hat{U} + \hat{H}_\mathrm{ph},
\end{equation}
and the d$\cdot$E Hamiltonian is given by
\begin{equation}
    \hat{H}_\mathrm{d\cdot E}=\hat{H}_\mathrm{M} + \hat{U}\hat{H}_\mathrm{ph}\hat{U}^\dagger.
\end{equation}

After transformation and binning, we get
\begin{align}\label{eq:dE}
    \hat{H}_{d\cdot E}&=\hat{H}_\mathrm{M} + \hat{H}_\mathrm{ph}^\mathrm{rel}+\sum_{n,\lambda}\hbar\omega_n (\hat{B}_{n,\lambda}^\dagger\hat{B}_{n,\lambda} + \frac{1}{2}) \nonumber + \sum_{j,n,\lambda} i\omega_n \sqrt{\frac{\ell_n({\bf r})}{V}} \vec{\bf \mu}_j\cdot\vec{\bf A}_n (e^{-i\bf\vec{q}\cdot\overline{r}_j} \hat{B}_{n,\lambda}^\dagger - e^{i\bf\vec{q}\cdot\overline{r}_j} \hat{B}_{n,\lambda}) \\ 
    &+ \sum_{j,k,n,\lambda} \frac{\omega_n\ell_n({\bf r})}{2 \hbar V} (\vec{\bf \mu}_j\cdot\vec{\bf A}_n)(\vec{\bf \mu}_k\cdot\vec{\bf A}_n) \left(e^{-i{\bf q}\cdot (\overline{\bf r}_j - \overline{\bf r}_k)} +e^{i{\bf q}\cdot (\overline{\bf r}_j - \overline{\bf r}_k)}\right)
\end{align}
where $\vec{\bf A}_n = \sqrt{\frac{\hbar}{2 \omega_n \epsilon_0}} \hat{\epsilon}_n$. Here, it is clear that the result is the regular $d\cdot E$ Hamiltonian but with the vector potential scaled by the enhancement factor $\ell_n({\bf r})$.

The Pauli-Fierz Hamiltonian can be easily derived by transforming Equation \ref{eq:dE} under the unitary
\begin{equation}
    \hat{U}=e^{-i\frac{\pi}{2}\hat{B}_{n,\lambda}^\dagger\hat{B}_{n,\lambda}}
\end{equation}
which results in
\begin{align}
    \hat{H}_\mathrm{PF}&=\hat{H}_\mathrm{M} + \hat{H}_\mathrm{ph}^\mathrm{rel}+\sum_{n,\lambda}\hbar\omega_n (\hat{B}_{n,\lambda}^\dagger\hat{B}_{n,\lambda} + \frac{1}{2}) \nonumber + \sum_{j,n,\lambda} \omega_n \sqrt{\frac{\ell_n({\bf r})}{V}} \vec{\bf \mu}_j\cdot\vec{\bf A}_n (e^{-i\bf\vec{q}\cdot\overline{r}_j} \hat{B}_{n,\lambda}^\dagger + e^{i\bf\vec{q}\cdot\overline{r}_j} \hat{B}_{n,\lambda}) \\ 
    &+ \sum_{j,k,n,\lambda} \frac{\omega_n\ell_n({\bf r})}{2 \hbar V} (\vec{\bf \mu}_j\cdot\vec{\bf A}_n)(\vec{\bf \mu}_k\cdot\vec{\bf A}_n) \left(e^{-i{\bf q}\cdot (\overline{\bf r}_j - \overline{\bf r}_k)} +e^{i{\bf q}\cdot (\overline{\bf r}_j - \overline{\bf r}_k)}\right).
\end{align}

A long-wave approximation has already been made in deriving the $d\cdot E$ and Pauli-Fierz Hamiltonians, but a longer-wave approximation is often also made where it's assumed that ${\bf q}\cdot (\overline{\bf r}_j - \overline{\bf r}_k)\approx 0$ such that the exponentials go away, giving
\begin{align}
    \hat{H}_\mathrm{PF}&=\hat{H}_\mathrm{M} + \hat{H}_\mathrm{ph}^\mathrm{rel}+\sum_{n,\lambda}\hbar\omega_n (\hat{B}_{n,\lambda}^\dagger\hat{B}_{n,\lambda} + \frac{1}{2}) \nonumber + \sum_{j,n,\lambda} \omega_n \sqrt{\frac{\ell_n({\bf r})}{V}} \vec{\bf \mu}_j\cdot\vec{\bf A}_n (e^{-i\bf\vec{q}\cdot\overline{r}_j} \hat{B}_{n,\lambda}^\dagger + e^{i\bf\vec{q}\cdot\overline{r}_j} \hat{B}_{n,\lambda}) \\ 
    &~~~~~~~~~~~~~~~~~~+ \sum_{j,k,n,\lambda} \frac{\omega_n\ell_n({\bf r})}{\hbar V} (\vec{\bf \mu}_j\cdot\vec{\bf A}_n)(\vec{\bf \mu}_k\cdot\vec{\bf A}_n) .
\end{align}

{\footnotesize
\bibliography{references.bib}
}